\documentclass[aps,prl,twocolumn,groupedaddress]{revtex4}
\usepackage{graphicx}
\begin{document}

\title{Brane Gravitational Extension of \\
Dirac's "Extensible Model of the Electron"}
\author{Aharon Davidson}
\email[Email: ]{davidson@bgu.ac.il}
\author{Shimon Rubin}
\email[Email: ]{rubinsh@bgu.ac.il}
\affiliation{Physics Department, Ben-Gurion University, Beer-Sheva 84105, Israel}

\date{July 9, 2009}

\begin{abstract}
	A gravitational extension of Dirac's "Extensible model of the
	electron" is presented.
	The Dirac bubble, treated as a 3-dim electrically charged brane,
	is dynamically embedded within a 4-dim $Z_{2}$-symmetric
	Reissner-Nordstrom bulk.
	Crucial to our analysis is the gravitational extension of Dirac's
	brane variation prescription; its major effect is to induce a novel
	geometrically originated contribution to the energy-momentum
	tensor on the brane.
	In turn, the effective potential which governs the evolution of
	the bubble exhibits a global minimum, such that the size of
	the bubble stays finite (Planck scale) even at the limit where the
	mass approaches zero.
	This way, without fine-tuning, one avoids the problem so-called
	'classical radius of the electron'.
\end{abstract}

\maketitle

The source singularity associated with a charged point particle is
physically unacceptable.
Einstein, in an attempt to deal with this problem, being apparently
ready to trade his own general relativity for the so-called unimodular
gravity \cite{Einstein}, has raised the intriguing possibility that
'Gravitational fields play an essential part in the structure of the
elementary particles of matter'.
Dirac, on the other hand, ignoring gravity altogether, has viewed
the electron as a breathing bubble \cite{Dirac} in the electromagnetic
field 'with no constraints fixing its size and shape'.
In this paper, equipped with modern brane gravity \cite{RandallSundrum},
we re-examine the idea of a finite size electrically charged elementary
particle.
To be more specific, we hereby extend Dirac's "Extensible model of
the electron" general relativistically, treating the Dirac bubble as a
gravitating brane.

The idea of a finite size classical electron dates back to Abraham,
Lorentz, and Poincare \cite{Abraham}.
Being the first to introduce the (still) pleasing idea that the rest mass
$m_e$ of (say) the electron is mainly of electromagnetic origin, they
have attempted to remove the electromagnetic point singularity by
invoking a localized charge density.
Their naive model suffers however from a variety of severe drawbacks
already at the classical level.
For example,

\noindent (i) The classical radius of the resulting configuration
(using $c=1$ units)
\begin{equation}
	R_{cl}\simeq\frac{e^2}{m_e} \sim 10^{-13} cm ~,
	\label{Abraham}
\end{equation}
known as the classical radius of the electron, turns out to be several
orders of magnitude larger than the scale where quarks and leptons still
exhibit a point-like behavior, and in particular

\noindent (ii) The stabilization
mechanism of the localized electromagnetic charge distribution is totally
unknown.

Einstein, intrigued by the latter problem, and being presumably
frustrated by the Reissner-Nordstrom metric singularity, has speculated
\cite{Einstein} that ''In the interior of every corpuscle (meaning a charged
elementary particle) there subsists a negative pressure the fall of which
maintains the electrodynamic force in equilibrium''.
At the early sixties, the idea of the electron being an extended object
was revisited by Dirac.
The action principle underlying his ''Extensible model of the electron''
\cite{Dirac} consists of two parts, namely
\begin{equation}
       S_{D}=- \int \frac{1}{4}F^{\alpha\beta}
       F_{\alpha\beta} \sqrt{- G}~d^4 y+
       \int \sigma\sqrt{- g}~d^3 x ~,
       \label{SD}
\end{equation}
where $G_{\alpha\beta}(y)$ is the \emph{flat} 4-dim background metric,
and $g_{\mu\nu}(x)=G_{\alpha\beta}(y(x))y^\alpha_{~,\mu}y^\beta_{~,\nu}$
is the \emph{curved} 3-dim metric induced on the surface of the bubble.
The latter is parameterized by $y^\alpha (x^{\mu})$.
Assuming now that the electron
surface is conductive, so that the electromagnetic fields inside
the bubble seized to exist, the 4-dim integral involving the Maxwell
field $A_\alpha (y) $ is carried out solely over the space outside
the electron.
A positive surface tension $\sigma>0$ seems then mandatory in
order 'to prevent the electron from flying apart under the Coulomb
repulsion of it surface charge'.
As expected, the total energy $E(R)$, associated with a Dirac
configuration of radius $R$, turns out to be the sum of electrostatic
and surface tension pieces, namely
\begin{equation}
      E(R)=\frac{e^{2}}{2R}+\sigma R^{2} ~.
      \label{energy}
\end{equation}
The minimization of this expression with respect to $R$ leads to
\begin{equation}
      R_e =\left( \frac{e^{2}}{4\sigma }\right) ^{1/3} , \\
      \quad m_e=\frac{3 e^2}{4R_e} ~.
      \label{RM}
\end{equation}
Unfortunately, the classical radius of the electron problem has not
gone away, and one still lacks the limit where, for a given $e$, both
$m_e$ and $R_e$ can be arbitrarily small.
What one is really after, however, is a classic configuration
characterized by
\begin{equation}
     R_e \ll R_{cl}=\frac{e^2}{m_e} \quad
     \text{for} \quad m_e \ll \frac{|e|}{\sqrt{G_N}}~.
\end{equation}
Over the years the source singularity problem has triggered a variety
of classical and semi-classical models of the electron, both without
\cite{models} and with \cite{Gmodels} interactions with gravity.
Most of the latter are in fact modern brane gravitational extensions
of Dirac's model.

On the technical side, however, Dirac has made an important
observation which goes well beyond the specific model constructed.
Facing the \emph{in/out} asymmetry between the \emph{inner} and
the \emph{outer} bulk sections, Dirac has pointed out \cite{Dirac}
that a 'naive variation' of the action eq.(\ref{SD}) is not a linear
function of $\delta y^\alpha(x^{\mu})$.
To be specific, if one makes a variation $\delta y^\alpha (x^{\mu})$,
corresponding to the surface being pushed out a little, $\delta S_D$ will
not be minus the $\delta S_D$ for $-\delta y^\alpha (x^{\mu})$,
corresponding to the surface being pushed in a little, on account of the
field just outside the surface being different from the field just inside.
Dirac's conclusion was that the choice of $y^\alpha (x^{\mu})$ as the
canonical variables will not do. To bypass this field theoretical problem,
he has ingeniously introduced general curvilinear coordinates such that, in
the new coordinate system, to be referred to as the Dirac frame, the
location of the bubble does not change during the variation process
A Dirac style brane variation has been recently applied \cite{DG} to
modern brane gravity.

\begin{figure}[th]
	\includegraphics[scale=0.45]{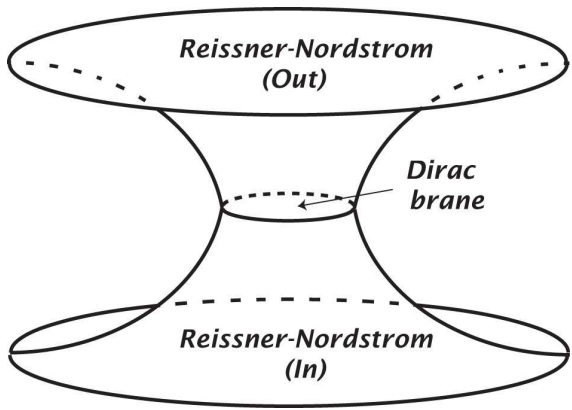}
	\caption{The Dirac brane is the source of a wormhole
	configuration gluing two Reissner-Nordstrom regimes.
	Unlike in the case of the Wheeler wormhole, the direction
	of the electric force lines are now opposite to each other
	in the two regimes.}
\label{DiracBrane}
\end{figure}

Dirac bubble separates two regions of space-time, to be
referred to as the \emph{in} and the \emph{out} regions in obvious
notations, such that the whole manifold is non-singular.
On convenience grounds, to make the calculations a bit simpler,
we impose, at a certain stage, a discrete \emph{ in-out}
$Z_{2}$ symmetry.
This way, the configuration resembles, in some sense, the
Einstein-Rosen bridge \cite{EinsteinRosen}, and the Wheeler-Misner
wormhole \cite{Wheeler}, but now, with the Dirac bubble serving as a
source matching its \emph{in} and \emph{out} regimes, see
Fig.(\ref{DiracBrane}).
It should be emphasized that by virtue of the Geroch-Traschen theorem
\cite{GerTra},
the Dirac brane, being a co-dimension 1 extended object, is a legitimate
source for the gravitational field [in contrast with a point-like source,
which is of co-dimension 3, and is not compatible with Einstein's
field equations as was demonstrated \cite{Paddy} for the particular
case of Schwarzschild spacetime].
In our notations, the bulk metric is given by $G_{\alpha \beta }^{i}(y)$,
where $i=in,out$ are separate coordinate systems associated with the
two regions, respectively.
Being the boundary of two regions, the location of the bubble is specified
by each of the embedding vectors $y^{i\alpha }(x)$, with $x^{\mu }$
spanning the local coordinate system on the bubble.
The induced metric on the bubble becomes
\begin{equation}
       g_{\mu \nu }(x)=
       G_{\alpha \beta }^{i}(y^{i}(x))y_{~,\mu }^{i\alpha }y_{~,\nu
       }^{i\beta }~.
\end{equation}
The underlying action principle is given by the following piecewise form
\begin{equation}
       S=S_{in}+S_{out}+S_{brane}+S_{\sigma }~.  \label{Action}
\end{equation}
The bulk actions $S_{i}$ in the various $i=in,out$ regions are given by
\begin{equation}
       S_{i}=-\int_{i}\left( \frac{R^{i}}{16\pi G_{N}}+
       \frac{1}{4}F_{i}^{\alpha
       \beta }F_{\alpha \beta }^{i}\right) \sqrt{-G^{i}}~d^{4}x~,
       \label{Setup}
\end{equation}
each of which being the Einstein-Hilbert action minimally coupled to an
electromagnetic field $A_{\alpha }^{i}(y)$.
In our notations,  $F_{\alpha \beta}^{i}=A_{\alpha ,\beta }^{i}(y)-
A_{\beta ,\alpha }^{i}(y)$.
The term $S_{brane}$ is itself a sum of three pieces
\begin{equation}
       S_{brane}=S_{K}^{in}+S_{K}^{out}+S_{constraint}~.  \label{Brane action}
       \end{equation}
The two accompanying extrinsic curvature terms
\begin{equation}
       S_{K}^{i}=\frac{1}{8\pi G_{N}}\int\limits_{brane}K^{i}\sqrt{-g}~d^{3}x
       \label{KTerm}
\end{equation}
are mandatory in a Lagrangian formalism, and are introduced to cancel
the non-integrable contributions arising from the variations of the
bulk metric.
The extrinsic curvature scalar $K$ denotes the trace of the extrinsic
curvature tensor $K_{\alpha \beta }$
\begin{equation}
       K=g^{\alpha \beta }K_{\alpha \beta } ~,
\end{equation}
and can be expressed in terms of the embedding vector and the normal to the
brane via
\begin{equation}
      K_{\alpha \beta }=\frac{1}{2}\left( n_{\alpha ,\mu }y_{,\nu }^{\alpha
      }+n_{\alpha ,\nu }y_{,\mu }^{\alpha }-2n_{\alpha }\Gamma _{\beta \gamma
      }^{\alpha }y_{,\mu }^{\beta }y_{,\nu }^{\gamma }\right)~.
      \label{Extrinsic}
\end{equation}
The $S_{K}^{i}$ terms are accompanied by the term $S_{constraint}$
which is given explicitly by \cite{DG}
\begin{equation}
   \begin{array}{c}
      S_{constraint}=\sum\limits_{i=in,out}\int \{\lambda _{i}^{\mu \nu }\left(
      g_{\mu \nu }-G_{\alpha \beta }^{i}y_{,\mu }^{i\alpha }y_{,\nu }^{i\beta
      }\right) + \vspace{4pt}\\
     + \eta _{i}^{\mu }y_{,\mu }^{i\alpha }n_{\alpha }^{i}+\chi _{i}(G_{\alpha
      \beta }^{i}n^{i\alpha }n^{i\beta }-1)\}\sqrt{-g}~d^{3}x~.
   \end{array}
\end{equation}
The set $\lambda _{i}^{\mu \nu },$ $\eta _{i}^{\mu },$ $\chi _{i}$ of
Lagrange multipliers reflects the constraints imposed the induced metric
$g_{\mu \nu }$ and on the normals $n^{i\alpha }$.
It is by no means mandatory to include the constrains explicitly in the
Lagrangian, but this way, the derivation of the equations of motion gets
significantly simplified [see Ref.\cite{DK} for derivation of the conventional
Israel junction conditions with the above constraints explicitly imposed].
The surface tension term $S_{\sigma }$, solely representing the localized
source distribution which actually forms the brane, is given by
\begin{equation}
       S_{\sigma }=\int\limits_{brane}\sigma \sqrt{-g}~d^{3}x~.
       \label{BraneAction}
\end{equation}
One may expect the mass of the configuration to vanish
at the limit $\sigma \rightarrow 0$.

The variation of the action eq.(\ref{Action}) with respect to the brane
metric $g_{\mu \nu }$ leads to Israel junction conditions \cite{Israel}
\begin{equation}
	\begin{array}{c}
	\frac{1}{2}T^{\mu \nu }+
	\sum\limits_{i=in,out}\lambda _{i}^{\ast
	\mu \nu }- \\
	\displaystyle{\frac{1}{8\pi G_{N}}
	\sum\limits_{i=in,out}(K_{i}^{\mu \nu }-g^{\mu
	\nu }K_{i})=0},
	\end{array}
	\label{Variation Brane Metric}
\end{equation}
where an additional energy momentum tensor, namely
$\lambda _{in}^{\ast\mu\nu}+\lambda _{out}^{\ast\mu\nu}$, has
been induced on the brane.
The latter can be interpreted as the energy-momentum piece which
accounts for the constraint forces which keep the brane still during
the variation.
In our notations,
\begin{equation}
      \lambda _{i}^{\ast \mu \nu }=
      \frac{1}{16\pi G_{N}}K_{i}^{\mu \nu }-\lambda
      _{i}^{\mu \nu }~.
      \label{lambdastar}
\end{equation}
Varying the action with respect to the bulk metric, it should be
re-emphasized that on the brane, following Dirac's variation prescription,
the variation is solely due to general coordinate transformation \cite{DG}.
This in turn leads to
\begin{equation}
    \begin{array}{l}
       \lambda _{i}^{\ast \mu \nu }K^{i}_{\mu \nu }=0 ~, \vspace{4pt}\\
       \lambda _{i~~;\nu }^{\ast \mu \nu }=0~ .
    \end{array}
\label{Equations for Lambda star}
\end{equation}
It is worth noticing that eq.(\ref{Equations for Lambda star}) does admit
$\lambda _{i}^{\ast \mu \nu }=0$ as a special solution, which upon
substituting to eq.(\ref{Variation Brane Metric}), recovers the conventional
Israel junction conditions \cite{Israel}.
However, relevant to the current discussion are the generic
$\lambda _{i}^{\ast \mu \nu } \neq 0$ solutions.

We are after a solution of the field equations which is
(i) Spherically symmetric, and (ii) Static in the bulk.
Note, however, that the radius $a(\tau)$ of the spherical brane,
where the matching is supposed to take place, needs not be a
constant and may in fact evolve in time.
The cosmic time $\tau$ on the brane is to be defined soon.
The dynamics which governs the
Dirac brane is dictated by the Israel junction conditions eq.(\ref
{Variation Brane Metric}).
Applying the Birkhoff theorem to each of the \emph{in} and
\emph{out} bulk regions separately, the resultant solution is
the Reissner-Nordstrom metric.
Using the conventional Schwarzschild coordinate system,
the associated bulk line elements are then
\begin{equation}
       ds_{i}^{2}=-f_{i}(R) dT^{2}+\frac{dR^{2}}{f_{i}(R) }
       +R^{2}(d\theta^{2}+\sin ^{2}\theta d\varphi^{2})~.
       \label{Intervals}
\end{equation}
The functions $f_{i}(R)$, which are $i$-independent
by virtue of the imposed $Z_{2}$ symmetry, are the familiar
\begin{equation}
       f(R) =1-\frac{2G_{N}M}{R}+\frac{G_{N}e^{2}}{R^{2}} ~,
\end{equation}
within the range $a(\tau)\leq R<\infty$.
We find it useful to write the function $f(R)$ in the form
\begin{equation}
       f(R) =1-\frac{2\zeta}{\rho} +\frac{\zeta}{\rho^{2} }~,
       \label{RN classical radius}
\end{equation}
such that $\zeta =G_{N}M/R_{cl}$ is a
dimensionless parameter, and $\rho =R/R_{cl}$
is the corresponding dimensionless radial marker.
For a consistently (soon to be properly) minimized $M$,
$R_{cl}=e^{2}/M$ would be the classical
radius of the electron.
Thus, to practically bypass the 'classical radius of the electron'
problem, one focuses his interest in the case  where on the
brane $\rho_{b}\ll1$, for which
$\displaystyle{\frac{2\zeta}{\rho_{b}} \ll\frac{\zeta}{\rho_{b}^{2} }}$.
Having in mind the Einstein corpuscle \cite{Einstein}, that is letting
gravity set the size of an elementary particle, one may further expect
\begin{equation}
	\frac{2\zeta}{\rho_{b}} \ll
	\frac{\zeta}{\rho^{2}_{b}}\simeq 1 ~.
\end{equation}

The embedding vector which describes the Dirac brane is specified by
\begin{equation}
	\begin{array}{l}
		T=T(\tau)~, \\
		R=a(\tau) ~,\\
		\theta=\theta ~,\\
		\varphi =\varphi ~,
		\end{array}
	 \label{Embedding}
\end{equation}
where the set $\{\tau ,\theta ,\varphi \}$ constitutes the local coordinate
system on the brane.
By invoking a proper gauge, i.e. for a proper choice of the function
$T(\tau)$, the line element associated with the induced brane metric
takes the isotropic form
\begin{equation}
      ds_{brane}^{2} =
      -d\tau^{2}+a(\tau)^{2}( d\theta ^{2} +
      \sin^{2}\theta d\varphi ^{2}) ~.
      \label{Brane Metric}
\end{equation}

Spherical symmetry of the solution makes the off-diagonal components
of eq.(\ref{Variation Brane Metric}) vanish, and reduces the number of
the independent diagonal components.
Furthermore, supplemented by Codazzi integrability condition, the total
number of independent field equations is reduced to a single equation.
Thus, on convenience grounds, we focus on the $\tau\tau$
component of eq.(\ref{Variation Brane Metric}).
For the given embedding vector, the relevant extrinsic curvature component
(see eq.(\ref{Extrinsic}), which enters eq.(\ref{Variation Brane Metric}), is
found to be
\begin{equation}
	K_{i}^{\theta\theta}=
	\frac{2}{a}\sqrt{f_{i}(a)+\dot{a}^{2}}  ~,
\end{equation}
and from eq.(\ref{Equations for Lambda star}) we deduce that
\begin{equation}
	\lambda _{i}^{\ast \tau \tau} =
	\frac{\omega _{i}}{a^{2}\sqrt{f_{i}(a)+\dot{a}^{2}}}  ~.
\end{equation}
The two constants of integration $\omega _{i}$ are about to play
a major role in our model.
Their values (actually the combination $\omega_{in} +\omega_{out}$)
are soon to be fixed on physical grounds.
The $\tau\tau$ component of eq.(\ref{Variation Brane Metric}) gives
rise to an implicit differential equation for the radius $a(\tau )$ of the
bubble, namely
\begin{equation}
   \begin{array}{c}
       \displaystyle{\frac{1}{2\pi G_{N}a}
       \left( \sqrt{f_{in}\left( a\right) +\dot{a}^{2}}+
       \sqrt{f_{out}\left( a\right) +\dot{a}^{2}}\right)+}\vspace{4pt} \\
       +2(\displaystyle{\lambda _{in}^{\ast \tau \tau }+
       \lambda _{out}^{\ast \tau \tau })=\sigma}~.
    \end{array}
\label{MechPotential}
\end{equation}

By algebraically solving for $\dot{a}^{2}$, the latter equation can
be re-arranged into
\begin{equation}
     \dot{a}^{2}+U_{mech}(a;M)=0~.
     \label{Potential}
\end{equation}
In this form, one actually faces an equivalent $1$-dim non-relativistic
mechanical problem with a vanishing total mechanical energy.
The family of mechanical potentials $U_{mech}(a;M)$ is parameterized
by the constant of integration $M$, and is explicitly given by
\begin{equation}
	\begin{array}{c}
	\displaystyle{U_{mech}(a;M)=
	1-\frac{2G_{N}M}{a}+\frac{G_{N}e^{2}}{a^{2}}}\vspace{4pt} \\
	\displaystyle{-\frac{1}{4}\pi ^{2}G_{N}^{2}a^{2}
	\left(\sigma-\sqrt{\sigma^{2}-\frac{8\omega}{\pi G_{N}
	a^{3}}}\right) ^{2}}~,
	\end{array}
	\label{Ueff}
\end{equation}
with $\omega =2(\omega _{in}+\omega _{out})$.
For $\omega \leq 0$, as hereby assumed, the mechanical potential
is well defined for all $a(\tau)>0$; it approaches unity when the radius
of the bubble tends to infinity, and diverges like $a^{-2}$ as the radius
tends to zero.
[Note in passing that the other solution, with a plus sign in front
of the square root, exhibits a wrong behavior at the large $a$ limit,
that is $U_{mech}\rightarrow -\infty$, and is thus physically unacceptable].

The associated static configuration, on which we focus attention,
calls for
\begin{equation}
	U_{mech}=0~, ~~~ \frac{dU_{mech}}{da}=0 ~.
	\label{UmechEqs}
\end{equation}
The combined solution of theses equations determines in turn
the mass $M_{0}$, as well as the radius $a_{0}$, of the static
configuration.
We now argue that, in contrast with the extensible Dirac model,
one can
(i) Fix the constant of integration $\omega$ such that the limit
$M_{0}\rightarrow 0$ becomes attainable, and
(ii) Do so while keeping the corresponding $a_{0}$ finite.
The fact that $a_{0}$ does not explode for a vanishing mass is
clearly the necessary ingredient for avoiding the 'classical radius
of the electron' problem.

\begin{figure}[th]
	\includegraphics[scale=0.45]{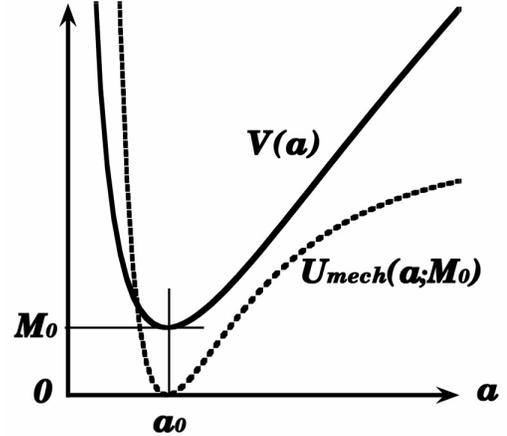}
	\caption{Owing to the implicit function theorem, the minimum
	$M_{0}$ of the function $V(a)$ (solid line) and the vanishing minimum
	of $U_{mech}(a;M_{0})$ (dashed line) occur at the same $a_{0}$.}
	\label{UmechV}
\end{figure}

Solving eq.(\ref{Potential}) for $M$, and treating it as the total
(conserved) energy $E(a,\dot{a})$ of the Dirac bubble, it can be
expressed in the form
\begin{equation}
	E(a,\dot{a})=\frac{a\dot{a}^{2}}{2G_{N}}+V(a)~,
\end{equation}
where the potential function is given by
\begin{equation}
	V(a)=\frac{e^{2}}{2a}+\frac{a }{2G_{N}}-
      \frac{G_{N}\pi ^{2}a^{3}}{8}\left( \sigma-
       \sqrt{\sigma^{2}-\frac{8\omega }{\pi G_{N}a^{3}}}\right) ^{2}~.
\end{equation}
Owing to the implicit function theorem, the minimum $M_{0}$
of the function $V(a)$ and the vanishing minimum of $U_{mech}(a;M_{0})$
occur at the same $a_{0}$, and consequently
eqs.(\ref{UmechEqs}) are translated into $V^{\prime}(a)=0$, see
Fig.(\ref{UmechV}).

The role of the (so far undetermined) negative constant of integration
$\omega$, contributing a negative definite term to $V(a)$, appears
to be crucial at small distances where
\begin{equation}
	V(a) \simeq \frac{e^{2}}{2a}-
	\pi |\omega|+\frac{a }{2G_{N}}
	\quad (\text{for}~ a\rightarrow 0)~,
	\label{smalla}
\end{equation}
and quite negligible at large distances, where
\begin{equation}
	V(a)\simeq \frac{a }{2G_{N}}+\frac{e^{2}}{2a}-
	\frac{2\omega^{2}}{G_{N}\sigma^{2}a^{3}}
	\quad (\text{for}~ a\rightarrow \infty)~.
\end{equation}
Furthermore notice that the leading terms in the small surface
tension expansion happen to coincide with eq.(\ref{smalla}).
This in turn calls for the physical choice
\begin{equation}
	\omega=-\frac{|e|}{\pi\sqrt{G_{N}}}  ~,
\end{equation}
which ensures that the $M_{0}\rightarrow 0$ limit be approached
in the $\sigma \rightarrow 0$ no-brane case.

\begin{figure}[th]
	\includegraphics[scale=0.45]{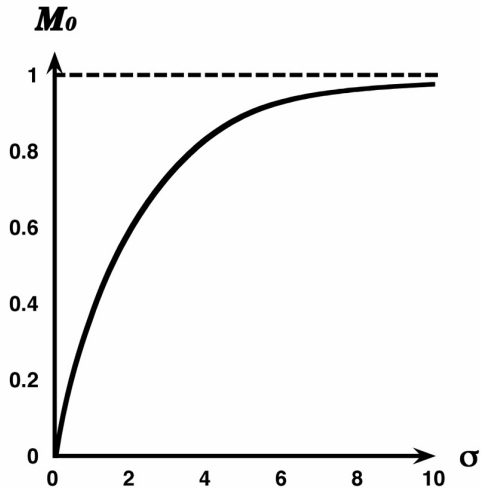}
	\caption{The mass of the gravitational Dirac bubble
	$M_{0}$ is plotted as a function of the surface tension
	$\sigma$ (in natural units specified by eq.(\ref{units})).
	The limit $M_{0}\rightarrow 0$ is associated with
	$\sigma\rightarrow 0$, whereas
	$M_{0}\rightarrow |e|/\sqrt{G_{N}}$ as
	$\sigma\rightarrow \infty$.}
	\label{mass}
\end{figure}

\begin{figure}[th]
	\includegraphics[scale=0.45]{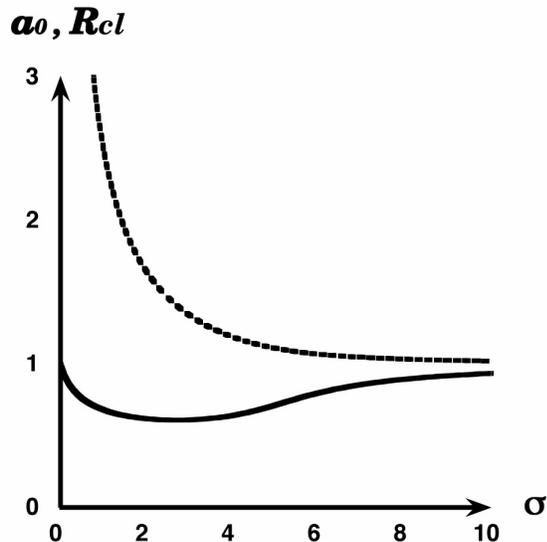}
	\caption{The radius of the gravitational Dirac bubble
	$a_{0}$ (solid line) is compared with the corresponding
	classical radius $R_{cl}$ (dashed line). Both radii are
	plotted as a function of $\sigma$.
	The natural units used are specified by eq.(\ref{units}).}
\label{RclR}
\end{figure}

Altogether, the result closely resembles a classical elementary
particle.
The mass $M_{0} (\sigma)\leq |e|/\sqrt{G_{N}}$ of the bubble
(the $M_{0} > |e|/\sqrt{G_{N}}$ case, which is perfectly allowed
by the Reissner-Nordstrom geometry, has no  representation here)
is plotted in Fig.(\ref{mass}) as a function of the surface tension.
Respectively, the radius $a_{0}(\sigma)$ of the bubble is plotted
in Fig.(\ref{RclR}), and is compared with the associated classical
radius $R_{cl}(\sigma)$.
The units used in these figures are natural, solely set by $e$
and $G_{N}$.
To be specific,
\begin{equation}
	M=\frac{|e|}{\sqrt{G_{N}}}\widetilde{M} ~,~
	a=\sqrt{G_{N}}|e|\widetilde{a} ~,~
	\sigma=\frac{\widetilde{\sigma}}{\pi G_{N}^{3/2} |e|} ~,
	\label{units}
\end{equation}
with $\widetilde{M},\widetilde{a},\widetilde{\sigma}$ being
dimensionless quantities.

\noindent $\bullet$ For small values of $\sigma$, that is for
$\widetilde{\sigma}\ll1$, one finds
\begin{equation}
	V(a) \simeq \frac{1}{2a}\left(|e|-\frac{a}{G_{N}}\right)^{2}
	+ \pi \sigma\sqrt{\frac{G_{N}^{1/2}|e|}{2}}a^{3/2} ~.
\end{equation}
This in turn describes a charged Dirac bubble of arbitrarily
small mass
\begin{equation}
	M_{0}=\frac{\pi  G_{N}e^2 \sigma}{\sqrt{2}}\ll
	\frac{|e|}{\sqrt{G_{N}}}=\sqrt{\alpha}M_{Pl}~,
\end{equation}
and of a Planck length size
\begin{equation}
	a_{0}=\sqrt{G_{N}}|e| =\sqrt{\alpha}L_{Pl}~,
\end{equation}
with $\alpha \simeq 1/137$ denoting the fine structure constant.
The latter length is clearly negligible, as previously argued,
in comparison with the corresponding classical radius $R_{cl}$,
which is of order $e^{2}/M_{0}$, that is
\begin{equation}
	\frac{a_{0}}{R_{cl}}=\sqrt{2} \pi
	\sigma G_{N}^{3/2}  |e| \ll1 ~.
\end{equation}

\noindent $\bullet$ For large values of $\sigma$, on the other
hand, the mass tends asymptotically to its upper bound
\begin{equation}
	M_{0} \rightarrow \frac{|e|}{\sqrt{G_{N}}} ~,
\end{equation}
while the radius is again of the Planck scale,
\begin{equation}
	a_{0} \rightarrow \sqrt{G_{N}}|e| ~,
\end{equation}
but is now not too different from the corresponding classical radius.
Serendipitously, at the $\sigma \rightarrow \infty$ limit, one faces
two extremal Reissner-Nordstrom black holes glued together at
their common horizon.
[see e.g. \cite{Israel2} for glueing two Kerr regimes at the ring
singularity, and \cite{Guendelman} for glueing two Schwarzschild
regimes, using a light-like brane, at their common horizon].

Needless to say, the hereby presented gravitational extension of
Dirac's ''Extensible model of the electron'' is extensible as well.
It is still a classical model, which does not incorporate the notion
of spin, and which fully ignores the electro/nuclear gauge interactions.
The stability of the static configuration with respect to non-spherical
shape deformations is yet to be considered in another paper.
Still, on pedagogical grounds, the incorporation of (unified) brane
gravity allows us to
(i) Get rid, by construction, of the electro/gravitational source singularities,
(ii) Bypass the problem so-called 'the classical radius of the
electron', and
(iii) Construct elementary Einstein corpuscles with arbitrarily small mass
as compact objects of the Planck length size.

\acknowledgments{It is our pleasure to cordially thank our colleague
Ilya Gurwich for valuable and inspiring remarks. We also thank Eduardo
Guendelman and Idan Shilon for useful discussions.}


\begin{thebibliography}{99}
\bibitem{Einstein}
	A. Einstein, Siz. Preuss. Acad. Scis. (part 1), 349 (1919), in \textit{The
	Principle of Relativity} (Dover, New York, 1923).
\bibitem{Dirac}
	P.A.M. Dirac, Proc. Roy. Soc. of London \textbf{A268}, 57 (1962);
	P. A. M. Dirac, C. Moller, and A. Lichnerowicz, Proc. Roy. Soc. of
	London \textbf{A270}, 354 (1962).
\bibitem{RandallSundrum}
	L. Randall and R. Sundrum, Phys. Rev. Lett. \textbf{83}, 3370 (1999);
	L. Randall and R. Sundrum, Phys. Rev. Lett. \textbf{83}, 4690 (1999);
	T. Shiromizu, K. Maeda and M. Sasaki, Phys. Rev. \textbf{D62},
	024012 (2000);
	F. Quevedo, Class. Quant. Grav. \textbf{19}, 5721 (2002);
	E. Papantonopoulos, Lect. Notes Phys. \textbf{592}, 458 (2002);
	D. Langlois, Prog. Theor. Phys. Supp. \textbf{148}, 181 (2003);
	P. Brax and C. Van de Bruck, Class. Quant. Grav. \textbf{20}, R201
	(2003);
	R. Maartens, Living Rev. Rel. \textbf{7}, 7 (2004);
	P. D. Mannheim, \textit{Brane-Localized Gravity} (World Scientific,
	Singapore, 2005).
\bibitem{Abraham}
	M. Abraham Phys. Zeit. \textbf{4}, 57 (1902); ibid.  Gottinger
	Nachrichten 20 (1902);
	H.A. Lorentz, in \textit{Theory of Electrons} 2nd ed. (1915),
	(Dover Publications, New York, 1952);
	H. Poincare, C. R. Acad. Sci. \textbf{140}, 1504 (1905).
\bibitem{models}
	H.B.G. Casimir, Physica XIX, 846 (1953);
	T.H. Boyer, Phys. Rev. \textbf{174}, 1764 (1968);
	P. Gnadig, Z. Kunszt, P. Hazenfratz and J. Kuti, Annals Phys.
	\textbf{116}, 380 (1978);
	T.H. Boyer, Phys. Rev. \textbf{D25}, 3246 (1982);
	A.O. Barut and N. Zanghi, Phys. Rev. Lett. \textbf{52}, 2009 (1984);
	A. Davidson and U. Paz, Phys. Lett. \textbf{B300}, 234 (1993);
	M. Pavsic, E. Recami, W.A. Rodrigues Jr., G.D. Maccarrone, F. Raciti
	and G. Salesi, Phys.Lett. \textbf{B318}, 481 (1993).
	A. O. Barut and M. Pavsic, Phys. Lett. B306, \textbf{49}, (1993);
\bibitem{Gmodels}
	A.V. Vilenkin and P.I. Fomin, Nuovo Cim. \textbf{A45}, 59 (1978);
	C.O. Lopez, Phys. Rev. \textbf{D30}, 313 (1984);
	O. Gron, Phys. Rev. \textbf{D31}, 2129, (1985);
	M. Onder and  R.W. Tucker, Phys. Lett. \textbf{202B}, 501 (1988);
	D.H Hartley, M. Onder and R.W. Tucker, Class. Quant. Grav. \textbf{6}
	1301 (1989);
	S. Ansoldi, A. Aurilia, R. Balbinot, and E. Spallucci, Phys. Essays
	\textbf{9}, 556 (1996);
	A. L. Larsen and C.O. Lousto, Nucl. Phys. \textbf{B472}, 361 (1996);
	O.B. Zaslavskii, Phys. Rev. \textbf{D70}, 104017 (2004).
\bibitem{DG}
	A. Davidson and I. Gurwich, Phys. Rev. \textbf{D74}, 044023 (2006);
	G. Kofinas and T. Tomaras, Class. Quant. Grav. \textbf{24}, 5861 (2007).
\bibitem{EinsteinRosen}
	A. Einstein and N. Rosen, Phys. Rev. \textbf{48}, 73 (1935).
\bibitem{Wheeler}
	C.W. Misner and J.A. Wheeler, Ann. Phys. \textbf{2}, 525 (1957);
	M. Visser, in \textit{Lorentzian Wormholes: From Einstein to Hawking}
	(Springer, Berlin, 1996).
\bibitem{GerTra}
	R. Geroch and J.H. Traschen, Phys. Rev. \textbf{D36}, 1017 (1987).
\bibitem{Paddy}
	J.V. Narlikar and T. Padmanabhan, Found. of Phys. \textbf{18} , 659  (1988);
	T. Ortin in \textit{Gravity and Strings}, sec. 7.2, (Cambridge Monographs,
	2004).
\bibitem{DK}
	A. Davidson and D. Karasik, Phys. Rev. \textbf{D60}, 45002 (1999).
\bibitem{Israel}
	K. Lanczos, Phys. Zeils. 23, 539 (1922); ibid. Ann. der Phys. 74, 518 (1924);
	G. Darmois, Memorial des Sciences Mathematics XXV (Gauthier-Villars, Paris,
	1927);
	W. Israel, Nuovo Cimento \textbf{B44}, 1 (1966); erratum, ibid. \textbf{B48},
	463 (1967).
\bibitem{Israel2}
	W. Israel, Phys. Rev. \textbf{D2}, 4 (1970).
\bibitem{Guendelman}
	E.I. Guendelman, A. Kaganovich, E. Nissimov, and S. Pacheva,  Forschr. Phys.
	\textbf{57}, 566 (2009).
\end{thebibliography}
\end{document}